\documentstyle[prl,aps,multicol,epsf,epsfig]{revtex}
 
\begin{document} 

\tightenlines

\title{Generalized persistence exponents: an exactly soluble model}
\author{A. Baldassarri$^{1,2}$, J.P. Bouchaud$^1$, I. Dornic$^{1,3}$, and C.
Godr\`eche$^{1,4}$} 
\address{
$^1$Service de Physique de l'\'Etat
Condens\'e, CEA-Saclay, 91191 Gif sur Yvette, France\\
$^2$INFN Sezione di Roma, Universit\'a di Roma``La Sapienza'', 
Roma, Italy\\
$^3$Laboratoire de Physique de la Mati\`ere Condens\'ee, Universit\'e
de Nice, France\\
$^4$Laboratoire de Physique Th\'eorique et Mod\'elisation, Universit\'e de
Cergy-Pontoise, France\\ }

\maketitle

\begin{abstract}
It was recently realized that the persistence exponent appearing in
the 
dynamics of nonequilibrium systems is a special member of a continuously
varying family of exponents, describing generalized persistence properties. 
We propose and solve a simplified model of coarsening, where time intervals between spin flips are independent, and distributed according to a L\'evy law.
Both the limit
distribution of the mean magnetization and the generalized persistence exponents are
obtained exactly.

\end{abstract}

\pacs{02.50.Ey, 05.40.+j, 05.50+q, 05.70.Ln}

\begin{multicols}{2} 

\narrowtext

\def\d{{\rm d}} \def\e{{\rm e}} \def\sign{{\rm sign}}
\def\eps{\varepsilon} \def\im{{\rm Im}}
\def\tpn{{T_n^{+}}} \def\tmn{{T_n^{-}}} \def\tpt{{T^{+}_t}}
\def\tmt{{T^{-}_t}}

The surprise caused by the discovery of new nontrivial exponents in
the dynamics of simple nonequilibrium systems \cite{marcos95,ising94}
motivated a long series of works, mainly devoted to the search of
simple models or experimental situations, where the so called
persistence exponents could be computed or measured
\cite{dhp1et2,diffu96,majum96,ben,lee,exp97a,how,listref}. 
More recently,
the idea of {\it persistent large deviations} \cite{dg98} led to the
introduction of families of new nontrivial persistence exponents in,
e.g., the one dimensional Glauber-Ising chain at zero temperature, or
the simple diffusion equation. 
The probability of persistent
deviations, defined as the probability that the mean magnetization
$M_t=t^{-1} \int_0^t \d u\, \sigma(u)$ of the spin (or of the sign of the
diffusing field) at a given site was, for all previous times, greater than
some
level
$x$, where
$-1\le x\le 1$, was observed to decay algebraically at large times, with an
exponent
$\theta(x)$ continuously varying with $x$. 
When $x=1$, this probability is the usual persistence probability, since
imposing $M_t=1$ requires that the spin never flipped.
Hence
$\theta(1)=\theta$, the usual persistence exponent. 
Furthermore, the
distribution of
$M_t$ does not peak around zero for $t \to \infty$, but tends to a
nontrivial limit distribution on $[-1,1]$, singular at both ends as
$(1\mp x)^{\theta-1}$.
For coarsening systems, computing the exact value of $\theta$  turns out to be
a hard
problem, so one does not expect the computation of $\theta(x)$, or even
of the distribution of the mean magnetization, to be easily reachable.
The origin of the difficulty is that spins at different sites are strongly
correlated.

The aim of this letter is to present an extremely simplified version of the
coarsening models mentioned above, which allows for exact analytical
expressions
both of the limit distribution of the mean magnetization $M$, and of the
generalized persistence exponents $\theta(x)$.
Despite its simplicity, this model retains the essential features of the
coarsening process, in particular its non stationary properties, as will be
discussed below. 

In this model, which describes the dynamics of a single spin,
we assume that the time intervals between spin flips are independent.
It is indeed intuitively clear that, because of the
ever growing size of domains in coarsening systems (or of the diffusion length in the
diffusion equation), a spin at a given site can remain in the same direction
for
a very long time before a domain wall crosses this particular point and flips
the spin in the  reversed direction. 
By definition of the persistence exponent $\theta$, the time
$\tau$ before a spin is flipped is very broadly distributed, with a power law
tail decaying as $\tau^{-1-\theta}$ for large $\tau$. 
The simplest approximation is therefore to neglect the correlations between
the
different time intervals between flips, all assumed to be distributed with the
same density $p(\tau)$, decaying as $\tau^{-1-\theta}$. 

For simplicity, the distribution of time intervals
$p(\tau)$ is chosen to be a positive stable L\'evy distribution of index
$0<\theta<1$ denoted by $L^b_\theta(\tau)$. 
(The case $\theta > 1$ will be discussed below.)
Its Laplace transform reads $\hat L^b_\theta(s)=\exp(-b s^\theta)$,
where $b$ is the scale factor of the distribution, i.e. the typical values of
$\tau$ are of order $b^{1/\theta}$ \cite{gk}.
As is well known,
$L^b_\theta(\tau)$ decays asymptotically as $\tau^{-1-\theta}$ \cite{gk}. 
We will always suppose that 
$\sigma(t=0)=1$.
On the time axis, the process thus defined is a renewal process.

We have investigated
the statistics of the process, both {\it after $n$ sign changes}, or
{\it at time $t$}, with very similar results in the asymptotic regime. After $n$
sign changes, the time elapsed and the magnetization of the spin read
\begin{eqnarray}
t_n&=&\tau_1+\tau_2+\cdots+\tau_n,\\ S_n&\equiv& t_n M_n =
\tau_1-\tau_2+\cdots+(-)^{n-1}\tau_n,
\end{eqnarray}
while, at time $t$, they are given by
\begin{equation}
t=t_{N_t}+\lambda,\quad  S_t\equiv t M_t = S_{N_t}+(-)^{N_t}\lambda.
\end{equation}
In the first case, $n$ is given and $t_n$ is a random
variable, while in the second one, $t$ is given and $N_t$ is the
random variable equal to the largest $n$ for which $t_n\le t$. Finally
$\lambda$ is the length of time measured backwards from $t$ to the
last event. The corresponding distributions are defined
as
\begin{eqnarray}
P(n,x)&=&P\left(M_n={S_n}/{t_n}\ge x\right),\\
P(t,x)&=&P\left(M_t={S_t}/{t} \ge x\right).
\end{eqnarray}
For distributions which are peaked around their means at large
times, these quantities are referred to as the probabilities of large
deviations and are exponentially decreasing with $n$ or $t$
respectively. 
In the present
case, where $p(\tau)=L^b_\theta(\tau)$ is a positive L\'evy
distribution, we find the limit distribution
\begin{eqnarray}
\label{px1}
P(x) &=&\lim_{n\to\infty}P(n,x)=\lim_{t\to\infty}P(t,x),\\
\label{px2}
&=&\frac{1}{\pi\theta}\left [\frac{\pi}{2}-
\arctan\left(\frac{r\omega^{-\theta}+\cos\pi\theta}{\sin\pi\theta}\right)\right
].
\end{eqnarray}
where $\omega=(1-x)/(1+x)$ and $r=1$ (see below). 

\begin{figure}
\centerline{
\epsfysize=6cm
\epsfig{file=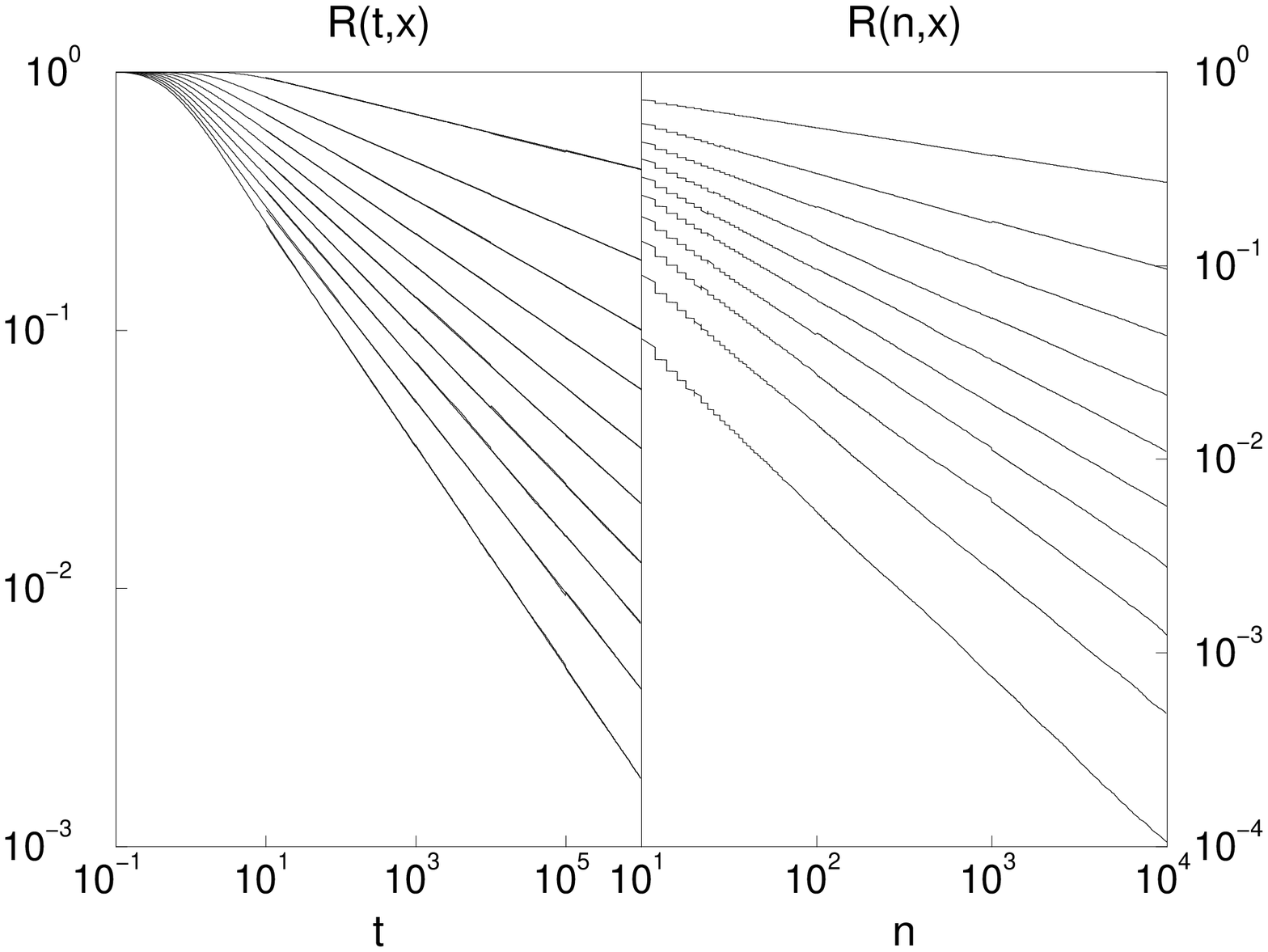,width=75mm,height=55mm,angle=0}}
\caption{\small Plot of $R(t,x)$ (left) and $R(n,x)$ (right) for
$\theta=1/2$ and various values of $x$, in log-log coordinates. The
power-law behavior of both quantities for large times is clearly
seen. 
\label{fig1} }
\end{figure}

Let us sketch the proof of (\ref{px2}) for $P(n,x)$,
leaving the calculation of $P(t,x)$ to a longer publication
\cite{ustocome}. 
We introduce $\tpn$ and $\tmn$, which are the lengths of time
spent by the spin, respectively in the positive or negative direction, such
that $t_n=\tpn+\tmn$ and $S_n=\tpn-\tmn$, with $\tpn=\tau_1+\tau_3+\cdots+\tau_{2k+1}$, if $n=2k+1$, and
$\tpn=\tau_1+\tau_3+\cdots+\tau_{2k-1}$, if $n=2k$, and
$\tmn=\tau_2+\tau_4+\cdots+\tau_{2k}$, in both cases. 
Then $P(S_n/t_n\ge x)=P(\tmn/\tpn \le \omega)$ with $\omega
=(1-x)/(1+x)$. 
Since $\tpn$ and $\tmn$ are sums of stable L\'evy
random variables $L^b_\theta$, they are themselves stable L\'evy
random variables $L^{b^{\pm}}_\theta$, where, using the addition rule
of the scale parameters, $b^-=kb$, and $b^+=kb$ (if $n=2k$), or
$b^+=(k+1)b$ (if $n=2k+1$). 
The determination of $P(n,x)$ therefore
amounts to computing the distribution of the ratio of two L\'evy laws
with parameters $b^-$ and $b^+$. Denoting by $H$ the Heaviside function,
and using its Laplace representation along the Bromwich contour, one finds
\begin{eqnarray}\nonumber
P(\tmn/\tpn &>& \omega)=\int_0^\infty\d\tau_1\d\tau_2
L^{b^+}_\theta(\tau_1) L^{b^-}_\theta(\tau_2)
H\left(\frac{\tau_2}{\tau_1}-\omega\right)\\ \nonumber
&=&\int\frac{\d s}{2i\pi s} \exp[-b^+
(s\omega)^\theta] \exp[-b^-(-s)^\theta].
\end{eqnarray}
This integral leads to (\ref{px2}) with $r=b^-/b^+$.
In the limit $n \to \infty$, $r \to 1$. This
derivation also shows that whenever $n$ is even, $P(n,x)=P(x)$. 

The limit density $f(x)=-P'(x)$ of the mean magnetization reads
\begin{equation}
\label{fx}
f(x)=\frac{\sin\pi\theta}{2\pi}\frac{2+\omega+\omega^{-1}}
{2\cos\pi\theta+\omega^{\theta}+\omega^{-\theta}}.
\end{equation}
It is even, and diverges when $x \to \pm1$ as $(1 \mp x)^{\theta-1}$.
For $\theta < \theta_c = 0.5946 \ldots$, where $\theta_c$ is
the solution of $\theta_c = \cos{(\pi \theta_c /2)}$, $x=0$
corresponds to a minimum of $f(x)$, while for larger $\theta$, it corresponds
to a local maximum.
This can be interpreted as a precursory sign of the fact
that $f(x)$ tends to $\delta(x)$ for $\theta > 1$.
(It also shows that $f(x)$ cannot be approximated by a beta
distribution when $\theta$ is too large. In this respect, compare to the
discussion in \cite{newman}.)

\begin{figure}[t]
\begin{center}
\epsfig{file=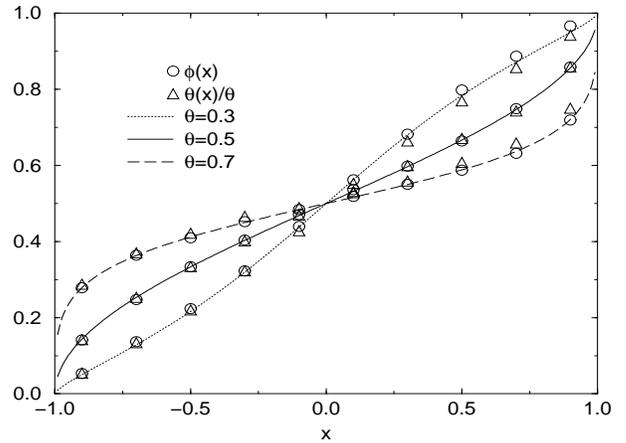,width=80mm,height=60mm,angle=0}
\caption{\small Plot of the exponents $\phi(x)$ and $\theta(x)/\theta$
  for $\theta=0.3,0.5,0.7$, showing
that the relation $\theta(x)=\theta \phi(x)$ holds. The lines
corresponds to the exact result $\phi(x)=1-P(x)$. \label{fig2} }
\end{center}
\end{figure}

We now consider the probability of persistent large deviations,
defined as the probability that the mean magnetization $M$ was,
for all previous times, greater than some level $x$. More precisely
one defines the quantities $R(n,x)=P(M_{n'}\ge x,\forall n'\le n)$
and similarly $R(t,x)=P(M_{t'}\ge x,\forall t'\le t)$.
Numerical computations show that both quantities decay algebraically in the
asymptotic regime (see Fig. 1), respectively as
$$
R(n,x)\sim n^{-\phi(x)}\ (n\gg1), \quad R(t,x)\sim t^{-\theta(x)}\ (t\gg1),
$$
where the two families of exponents are related by $\theta(x)=\theta
\phi(x)$ (see Fig. 2). 
This relation
is indeed expected since for a given $n$, $t_n$
scales as $n^{1/\theta}$. 
Note that by definition of the model,
$\theta(1)=\theta$. 
We also observe with very good accuracy (see
Fig. 2) the relation 
\begin{equation}
\label{phi}
\phi(x)=1-P(x)=\int_{-1}^x\d u f(u),
\end{equation}
which we now establish exactly.

For this, we note that $R(n,x)$ is the joint probability that $S_{n'}
\ge x t_{n'}$ for all $1 \leq n' \leq n$. 
Since clearly $R(2k,x)=R(2k+1,x)$, we
assume that $n$ is even, and write
\begin{eqnarray}\nonumber
R(n=2k,x) = P(&\xi_1& \ge 0, \xi_1 + \xi_2 \ge 0, \ldots, \\ &\xi_1& + \xi_2
+ \cdots + \xi_k \ge 0),
\label{andersen}
\end{eqnarray}
where $\xi_i=(1-x) \tau_{2i-1} - (1+x) \tau_{2i}$. 
Since the $\tau_i$
are positive L\'evy variables of index $\theta$, the $\xi_i$ are also
 L\'evy variables of index $\theta$, with an asymmetry parameter
$\beta=(\omega^{\theta}-1)/(\omega^{\theta}+1)$, which measures the
relative weight of the negative and positive tails \cite{gk}. 
The solution to (\ref{andersen}) for general stable L\'evy variables
is known \cite{andersen,feller}. 
It reads
\begin{equation}
R(n=2k,x) = \frac{\Gamma(k+1-q)}{\Gamma(k+1) \Gamma(1-q)},
\label{andersen2}
\end{equation}
where $1-q$ is the probability that $\xi$ is positive. 
This probability is precisely the
quantity $P(n=2,x)$ introduced above, itself equal to $P(x)$.
Hence $q=1-P(x)$. 
Finally, the large $k$ behavior of the r.h.s. of
(\ref{andersen2}) is  $\propto k^{-q}$, i.e., $\phi(x)=q$, which completes the
proof of (\ref{phi}).
Note that the plot of (\protect\ref{andersen2}) is indistinguishable from 
that obtained numerically for  $R(n,x)$. 
Eqs.  (\ref{px2}), (\ref{phi}), and (\ref{andersen2}) are the main results of this work.

The rest of this paper is devoted to a discussion of further properties of the model, and to a presentation of some 
possible generalizations. 

First, the stochastic process presented above, where time
intervals between spin flips are independent and distributed according to a
L\'evy distribution, exhibits nontrivial temporal properties, both from
mathematical
\cite{feller,gk}, and physical
\cite{trap1,derrida,bm} points of view. 
For example, although $p(\tau)$ is fixed in time, the probability distribution
of the length of time $\tilde \lambda$ from some time origin (or waiting time)
$t_{\rm w}$ to the next flip is non stationary for $\theta < 1$, i.e. it
depends
both on
$t_{\rm w}$ and 
$\tilde
\lambda$, while it is asymptotically independent of $t_{\rm w}$ for $\theta
>1$.
As a consequence, the probability that a given
spin did not flip between times $t_{\rm w}$ and $t_{\rm w}+t$ is a function of
$t_{\rm w}/t$ if $\theta<1$, while it is independent of $t_{\rm w}$ if
$\theta>1$ \cite{trap1}. 
Thus for $\theta < 1$, this model captures the {\it aging}
\cite{review} nature of the persistence phenomenon.
This property is deeply related to the fact that
the largest $\tau_i$ in the sum $t_n=\sum_{i=1}^n \tau_i$ contributes to
a finite fraction of $t_n$ for $\theta < 1$ even in the limit $n
\to \infty$, while this fraction is asymptotically zero for $\theta
\geq 1$ \cite{derrida,bm}.
Correspondingly, this also ensures that the distribution of
the mean magnetization does not peak around $x=0$, as was shown above. 

Despite its simplicity, the model discussed here thus shares many features of more
complex coarsening processes. 
As shown above, it leads to nontrivial predictions for the quantities $P(x)$
and $\theta(x)$.
Also, the behavior of $R(t,x)$ observed in Fig. 1 strongly resembles that
found in \cite{dg98} for the Glauber-Ising chain or the diffusion equation.
These predictions can be seen as approximations for these more general models.
In Fig. 3, we compare, for the Glauber model at zero temperature, the function
$\theta_{\rm{G}}(x)/\theta_{\rm{G}}(1)$,
as determined numerically in
\cite{dg98}, both with $1-P(x)$, where $P(x)$ is given by
(\ref{px2}) with $\theta=3/8$ \cite{dhp1et2}, and with
$1-P_{\rm{G}}(x)$, the distribution of magnetization measured numerically in
\cite{dg98}.  
Although there is qualitative agreement between the three curves,
the above relations are clearly only approximate. 
Furthermore there remains to understand the qualitative difference in
behavior between
the persistence exponents $\theta(x)$ for diffusion and for the Glauber-Ising
chain. It would be therefore interesting to generalize the present model to
include
some correlations between the time intervals $\tau_i$. The independent
interval model 
presented here is actually, in many respects, similar
to the random energy model ({\sc rem}) for spin-glasses \cite{derrida,bm}.
For example, the r.h.s. of (\ref{andersen2}) is identical to
the
expression for the participation ratio $Y_{k+1}$ in the {\sc rem}, with 
a reduced temperature equal to $q$ \cite{derrida,bm,ustocome}. 

\begin{figure}
\centerline{
\epsfig{file=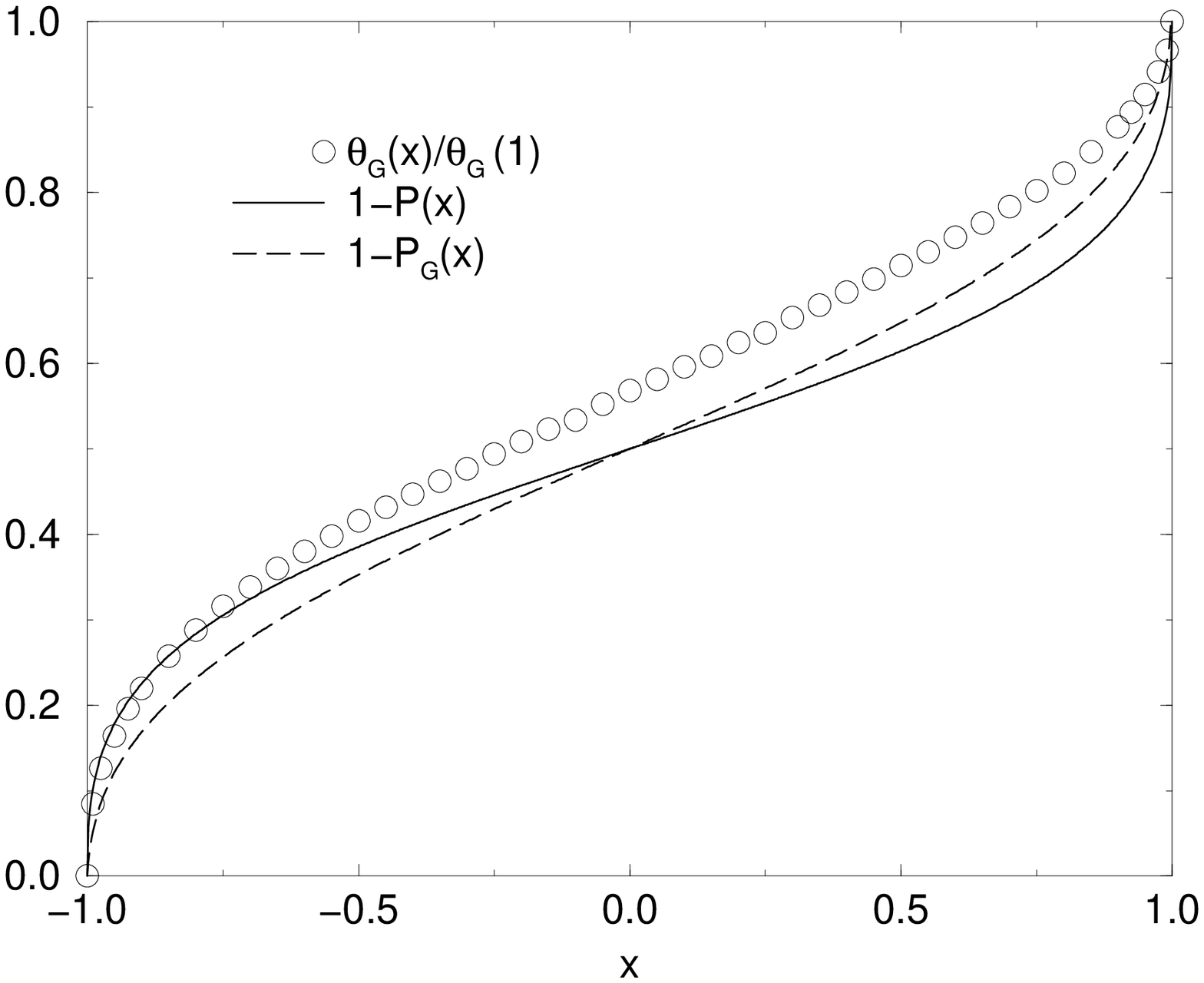,width=80mm,height=60mm,angle=0}} {\small
FIG.~3. Comparison between the function $\theta_{\rm{G}}(x)
/\theta_{\rm{G}}(1)$ for the Glauber-Ising model in one
dimension, as determined numerically in \protect\cite{dg98}, and two
predictions inspired from the present model: $1-P(x)$, where $P(x)$ is
given by  (\protect\ref{px2}) with $\theta=3/8$, and $1-P_{\rm{G}}(x)$
(see \protect\cite{dg98}). \label{fig3} }
\end{figure}

The only feature of $p(\tau)$ relevant for the results given
here is its asymptotic power-law behavior at large values of $n$ or $t$, 
and not its detailed behavior for small $\tau$.  
The reason is that the sum of
a large number of power-law distributed variables converges (for $\theta < 1$)
towards the L\'evy distributions considered here. 
In our simulations we used two different
distributions both decaying as $\tau^{-1-\theta}$ for large $\tau$, obtaining
the same results for sufficiently large times.  
Note that the
relations
$\phi(x)=\theta(x)/\theta=1-P(x)$ are best obeyed numerically for
$\theta=1/2$, because in this case it is easy to generate
the corresponding stable distribution. 

We also studied the case $\theta > 1$, where $p(\tau)$ has a
finite first moment. 
In this case, it is easy to check that $t_n$ grows linearly
with $n$, while $S_n$ grows as $n^{1/\theta}$ for $1 <\theta < 2$ and as
$\sqrt{n}$ for $\theta > 2$ \cite{gk}. 
Hence, the quantity $M_n$ tends to
zero for large $n$, and $f(x)$ collapses to a $\delta$ function. However, the
persistence exponents
$\theta(x)$ remain well defined, and are found to be equal to
$\theta(x>0)=\theta$, $\theta(x=0)=1/2$ and $\theta(x<0)=0$. 
This shows that the relation between $\theta(x)$ and $P(x)$ actually still
holds
in this degenerate case, except for $x=0$ where the value of $P(x)$ is ill defined. 
However, the nature of the persistence 
phenomenon 
in this model is quite different when $\theta > 1$, where it becomes stationary (see above). 
It would be interesting to
see if this is also true of more general models where $\theta > 1$,
such as the diffusion equation in high dimensions \cite{diffu96,dg98,newman}.

We have generalized slightly the problem, by choosing the
time intervals during which the spin $\sigma$ is respectively equal to 1 or
to $-1$ with a different scale factor. 
The distribution $f(x)$ becomes asymmetrical.
One can however check, both numerically and analytically, that
the relations $\theta(x)/\theta=\phi(x)=q=1-P(x)$ still hold in this
case. 
We have also relaxed the deterministic alternation of signs, and considered
$S_n=\sum_{i=1}^{n} a_i \tau_i$, where the $a_i$ are independent
identically distributed random variables, with $\langle a \rangle = 0$
and $\langle a^2 \rangle $ finite. 
The above results (\ref{px2}) and (\ref{phi}) for the limit distribution and the exponents  remain unchanged. 

When $\theta=1/2$,
$p(\tau)=L_{1/2}^b(\tau)$ is precisely the distribution of the time
intervals between two returns to the origin of the   binomial
random walk with equal steps $\pm1$, in the regime of long times. 
In this sense the binomial random walk is `primitive' with respect to the present
`walk' with time-space coordinates $t$, $S_t$ (or $t_n$, $S_n$),
and instantaneous velocity equal to $\sigma(t)$.  
For the latter, $y=\tpt/t$, is the fraction of time spent by
the walk stepping in the positive direction, or for the primitive walk, the
fraction of time spent on the positive half axis. 
Its distribution is well known, and given at large times by the
arc sine density $1/\pi\sqrt{y(1-y)}$, which is precisely the result
(\ref{fx}), with $\theta=1/2$, and $x=S_t/t=2\tpt/t-1$. 
In this respect, (\ref{fx}) appears as a generalization of the arc sine law.
A striking consequence of the present work is the existence of the
families of exponents $\theta(x)$ and $\phi(x)$, since, when $\theta=1/2$,
these
exponents describe properties of the simple random walk.
We note that the present work provides a clue to the determination 
of the temporal behavior of the hierarchy of
quantities introduced in \cite{dg98}.

Finally, it is tempting to conjecture that for a generic stochastic process
$S_t$ such that there exist an $\alpha > 0$ for which $S_t/t^{\alpha}$ admits
a
nontrivial limit distribution at large times, then
\begin{equation}
P\left(\frac{S_{t'}}{t'^\alpha} \geq x, \forall t' < t\right) \propto
t^{-\theta(x)}
\end{equation}
Such a conjecture, verified in our model, is also corroborated by the works of
\cite{breiman,redner} on the random walk (for which $\alpha=1/2$). 
In this case $x$ is not restricted to a finite interval, and
${\theta}(x)$ cannot be simply expressed in terms of the Gaussian
distribution of $S_{t}/\sqrt{t}$.

We wish to thank J.M. Luck, S. Redner, and C. Sire for interesting
discussions.

\end{multicols}

\end{document}